\documentclass{llncs}
\usepackage{amssymb}
\usepackage{graphics}
\usepackage{graphicx}

\usepackage{url}
\usepackage{breakurl} 
\usepackage[breaklinks]{hyperref}
\usepackage[table]{xcolor} 
\newcolumntype{?}{!{\vrule width 0.75pt}}
\usepackage{makecell}

\usepackage[T1]{fontenc} 
\usepackage[utf8]{inputenc}

\usepackage{listings}
\usepackage{paralist}
\usepackage{printlen}
\usepackage{acronym}
\usepackage{units}
\usepackage{times}

\usepackage{textcomp}

\usepackage{fixltx2e}

\usepackage{lipsum}

\usepackage[tracking=true]{microtype}

\usepackage{array, caption, tabularx, makecell, booktabs}
\usepackage{tikz}
\usetikzlibrary{arrows,positioning}

\usepackage[labelfont=bf]{caption}
\usepackage[labelformat=simple]{subcaption}

\usepackage{tikz}
\newcommand\encircle[1]{%
\tikz[baseline={([yshift=-8pt]current bounding box.north)}]
   \node (X) [draw, shape=circle, inner sep=0, fill=black, text=white,scale=0.7] {\strut #1};}

\acrodef{cdn}[CDN]{Content Delivery Network}
\acrodef{as}[AS]{Autonomous System}
\acrodef{isp}[ISP]{Internet Service Provider}
\acrodef{cdi}[CDI]{Content Distribution Infrastructure}
\acrodefplural{cdi}[CDIs]{Content Distribution Infrastructures}
\acrodef{rpi}[RPi]{Raspberry Pi}

\newcommand{\eg}{e.g., }

\newcommand{\ie}{i.e., }

\newcommand{\etal}{et~al.}

\usepackage{xcolor}

\newcommand{\afblock}[1]{\noindent{\textbf{#1 }}}
\newcommand{\takeaway}[1]{\noindent{\textbf{Takeaway.}} \textit{#1}}

\urldef{\mails}\path|{hohlfeld, rueth, wolsing, tzimmermann}@comsys.rwth-aachen.de|

\usepackage[export]{adjustbox}

\def\CDXfirstMostAppCount{{55}}

\def\CDXfirstMostAppsNumber{{1}}

\def\CDXsecondMostAppCount{{20}}

\def\CDXsecondMostAppsNumber{{2}}

\def\CDXLargestAppCount{{1}}

\def\CDXLargestAppsNumber{{84}}

\def\TTLRankAkamai{{1}}

\def\TTLRankChosenTimesAkamai{{32.9\%}}

\def\TTLRankNameAkamai{{Akamai}}

\def\TTLRankBoxFilepathAkamai{{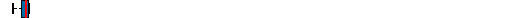}}

\def\TTLRankCDNetworks{{2}}

\def\TTLRankChosenTimesCDNetworks{{31.5\%}}

\def\TTLRankNameCDNetworks{{CDNetworks}}

\def\TTLRankBoxFilepathCDNetworks{{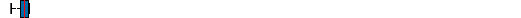}}

\def\TTLRankDNSDD{{3}}

\def\TTLRankChosenTimesDNSDD{{19.2\%}}

\def\TTLRankNameDNSDD{{DNSDD}}

\def\TTLRankBoxFilepathDNSDD{{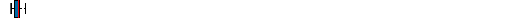}}

\def\TTLRankEdgecast{{4}}

\def\TTLRankChosenTimesEdgecast{{14.8\%}}

\def\TTLRankNameEdgecast{{Edgecast}}

\def\TTLRankBoxFilepathEdgecast{{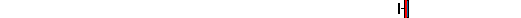}}

\def\TTLRankLevel3{{5}}

\def\TTLRankChosenTimesLevel3{{13.4\%}}

\def\TTLRankNameLevel3{{Level3}}

\def\TTLRankBoxFilepathLevel3{{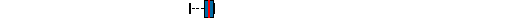}}

\def\TTLRankCloudflare{{6}}

\def\TTLRankChosenTimesCloudflare{{8.7\%}}

\def\TTLRankNameCloudflare{{Cloudflare}}

\def\TTLRankBoxFilepathCloudflare{{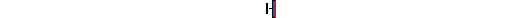}}

\def\TTLRankCDNWD{{7}}

\def\TTLRankChosenTimesCDNWD{{8.3\%}}

\def\TTLRankNameCDNWD{{CDNWD}}

\def\TTLRankBoxFilepathCDNWD{{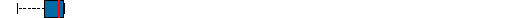}}

\def\TTLRankChinaCache{{8}}

\def\TTLRankChosenTimesChinaCache{{7.3\%}}

\def\TTLRankNameChinaCache{{ChinaCache}}

\def\TTLRankBoxFilepathChinaCache{{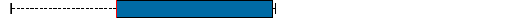}}

\def\TTLRankCloudfront{{9}}

\def\TTLRankChosenTimesCloudfront{{5.8\%}}

\def\TTLRankNameCloudfront{{Cloudfront}}

\def\TTLRankBoxFilepathCloudfront{{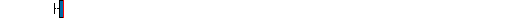}}

\def\TTLRankHighwinds{{10}}

\def\TTLRankChosenTimesHighwinds{{5.0\%}}

\def\TTLRankNameHighwinds{{Highwinds}}

\def\TTLRankBoxFilepathHighwinds{{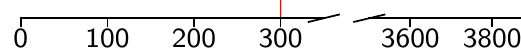}}

\newcommand\copyrighttext{%
  \footnotesize The final publications is available at Springer via \url{http://dx.doi.org/10.1007/978-3-319-76481-8_9}}
\newcommand\copyrightnotice{%
\begin{tikzpicture}[remember picture,overlay]
\node[anchor=south,yshift=10pt] at (current page.south) {\fbox{\parbox{\dimexpr\textwidth-\fboxsep-\fboxrule\relax}{\copyrighttext}}};
\end{tikzpicture}%
}

\begin{document}
\frontmatter          

\mainmatter  

\title{Better Cat Picture Delivery with Meta-CDNs}
\title{Characterizing a Meta-CDN}
\title{Always the best? Characterizing a Meta-CDN}
\title{A First Look on Meta-CDNs:\\ Architecture, Performance, and Implications}
\title{Characterizing a Meta-CDN}

\titlerunning{Characterizing a Meta-CDN}
\titlerunning{Always the best? Characterizing a Meta-CDN}
\titlerunning{A first look on Meta-CDNs: Architecture, Performance, and Implications}
\titlerunning{Characterizing a Meta-CDNs: Architecture, Methods, and Implications}
\titlerunning{Characterizing a Meta-CDN}

\author{Oliver Hohlfeld, Jan R\"uth, Konrad Wolsing, Torsten Zimmermann\textsuperscript{\textdagger} \\
{\small\textsuperscript{\textdagger}\textbf{Authors in alphabetical order}}}

\authorrunning{O. Hohlfeld, J. R\"uth, K. Wolsing, and T. Zimmermann}

\institute{Communication and Distributed Systems, RWTH Aachen University\\
\mails
}

\toctitle{Lecture Notes in Computer Science}
\tocauthor{Authors' Instructions}

\maketitle
\copyrightnotice
\begin{abstract}
CDNs have reshaped the Internet architecture at large. 
They operate (globally) distributed networks of servers to reduce latencies as well as to increase availability for content and to handle large traffic bursts. 
Traditionally, content providers were mostly limited to a single CDN operator.
However, in recent years, more and more content providers employ multiple CDNs to serve the same content and provide the same services.
Thus, switching between CDNs, which can be beneficial to reduce costs or to select CDNs by optimal performance in different geographic regions or to overcome CDN-specific outages, becomes an important task. 
Services that tackle this task emerged, also known as CDN broker, Multi-CDN selectors, or Meta-CDNs.
Despite their existence, little is known about Meta-CDN operation in the wild. 
In this paper, we thus shed light on this topic by dissecting a major Meta-CDN. 
Our analysis provides insights into its infrastructure, its operation in practice, and its usage by Internet sites. 
We leverage PlanetLab and Ripe Atlas as distributed infrastructures to study how a Meta-CDN impacts the web latency.

\end{abstract}

\section{Introduction}
\acp{cdn} have become a key component of the web~\cite{adhikari2012tale,calder2015anycast}.
Their ongoing quest to serve web content from nearby servers has flattened the hierarchical structure of the Internet~\cite{labovitz2010internet} and promises lower latencies, while their distributed nature promises high availability.
These benefits led to a wide adoption of \acp{cdn} for web content delivery that is manifested in high traffic shares: for example, more than half of the traffic of a North American~\cite{gerber2011backbone} or a European~\cite{poese2010padis} \ac{isp} can be attributed to few \ac{cdn}s only.
Despite these benefits, customers of a single \ac{cdn} are bound to its cost model and performance figures---a limitation that is solved by multihoming content on different \acp{cdn} and subsequently serving it from the \ac{cdn} that currently offers better performance and/or lower costs.

To better utilize content-multihoming, {\em Meta-CDNs}~\cite{frank2013pushing} enable content providers to realize custom and dynamic routing policies to direct traffic to the different \acp{cdn} hosting their content;
A concept also known as CDN-Selector~~\cite{xue2017cdn} and that is related to auction-based CDN brokers~\cite{mukerjee2016broker,BruceBroker17}.
Request routing is performed by the Meta-CDN according to {\em custom routing logic} defined by content-providers (i.e., the customers of a Meta-CDN and \acp{cdn}).
This routing logic can be informed by a broad range of factors, including \ac{cdn} cost models or measured \ac{cdn} performance.
Content providers can thus utilize a Meta-CDN to reduce costs or to optimize performance, \eg by implementing custom logic to direct traffic to a \ac{cdn} that currently offers better performance and/or lower cost (\eg at certain geographic regions or times).
Since the routing approach employed by the Meta-CDN customers is unknown to the involved \acp{cdn}, directed traffic and thus generated revenue gets harder to predict.
In particular, since decisions can be based on active performance measurements by the Meta-CDN, a (single) delivery of bad performance by the probed \ac{cdn} can result in rerouting traffic to a competing \ac{cdn} and thus losing revenue.
Thus, while Meta-CDNs can offer cost and performance benefits to content providers, they also challenge \ac{cdn} business models.
Concerning Internet-users, performance-based routing decisions can yield better Internet performance and benefit end-users while cost-based decisions can have other effects (as for any server selection approach run by \acp{cdn}).
While the concept is known and related work covering service specific implementations, \ie Conviva's streaming platform~\cite{conviva,dobrian2011conviva,mukerjee2016broker}, exists, the empirical understanding of a generic Meta-CDN and its operation in practice is still limited.
We posit that this understanding is necessary.

In this paper, we thus shed light on the Meta-CDN operation by dissecting the Cedexis Meta-CDN as a prominent example that is used by major Internet companies such as Microsoft (Windows Update and parts of the XBox Live Network), Air France, and LinkedIn~\cite{cedexis}.
Given its current adoption, understanding its functionality and its usage by customers provides a first step towards understanding currently unknown implications of Meta-CDNs on Internet operation.
We thus investigate the infrastructure and services powering this Meta-CDN and provide insights about its operation in practice.
We analyze for \emph{what} kind of services, \eg media, API backends, or bulk data transfers, customers utilize Cedexis and \emph{how} different \acp{cdn} are employed.
We further investigate how the infrastructure deployed by Cedexis impacts the overall request latency performance in a PlanetLab and Ripe Atlas measurement.
Specifically, our contributions are as follows:
\begin{enumerate}[i)]\vspace{-0.5em}
\item We characterize Cedexis, as a representative generic Meta-\ac{cdn}, present its operation principles, and further analyze and classify its customer base. Moreover, we illustrate which \acp{cdn} are used by the customers. 
\item We utilize globally distributed vantage points, \ie Ripe Atlas, PlanetLab, Open DNS Resolvers and a small deployment of probes behind home user DSL connections, to obtain a \emph{global} view on Cedexis. 
Based on these measurements, we analyze the deployed infrastructure of Cedexis, and are further able to investigate if the selection process varies based on the location.
In addition, we find cases of suboptimal routing in terms of latency.
\end{enumerate}

\vspace{-2em}
\section{Background and Related Work}
 \label{sec:rw}
\vspace{-1em}
To achieve high availability, content and service providers typically employ \ac{cdn} operators and utilize their already deployed and geographically distributed infrastructures~\cite{adhikari2012tale,calder2015anycast}.
In addition to increased availability, end-users profit from the distributed nature of these \acp{cdn} when retrieving content from close-by servers, reducing the overall latency.

Multiple works from academia and industry have investigated these infrastructures, the operation principles as well as the performance of deployed \acp{cdn}~\cite{adhikari2012tale,calder2015anycast,nygren2010akamai,otto2012cdn}.
Besides understanding and measuring \ac{cdn} infrastructures, researchers have utilized \ac{cdn} routing techniques to derive network conditions~\cite{su2009drafting}.
In addition, approaches that optimize the routing of user requests to the respective servers within a \ac{cdn}, as well as, optimized anycast load balancing have been proposed~\cite{chen2015mapping,flavel2015fastroute}.
Poese \etal~\cite{poese2010padis} present and analyze the impact of an ISP recommendation service providing insights about the current network state, \eg topology, load, or delay, to the \ac{cdn}, which in turn bases its server selection on the returned information. 
The idea and concept of the presented approach is revisited by Frank \etal~\cite{frank2013pushing} and, among other features, enables a \ac{cdn} to allocate server resources within the \ac{isp} on-demand when necessary.

To further ensure the availability of content, customers may use multiple \ac{cdn} deployments.
Other reasons to utilize more than one \ac{cdn} provider may be cost efficiency, \eg different prices to serve content at different times or due to traffic volume contracts.
However, with multiple locations at different \acp{cdn} serving the same service or content, either the customer or an additional service has to choose between the actual \ac{cdn} when a user requests a service or content~\cite{frank2013pushing,liu2012optimizing,xue2017cdn}.
With respect to \emph{multi-homed} content, \ie content that is distributed by multiple \acp{cdn}, Lui \etal~\cite{liu2012optimizing} present one of the first frameworks optimizing performance and cost of the resulting \ac{cdn} assignment.
In the case of video streaming, Conviva~\cite{conviva} uses a recommendation system that is utilized by the video player software~\cite{dobrian2011conviva} for the \ac{cdn} selection. 
Besides Conviva, commercial solutions that offer to act as the \ac{cdn} selector in more generic settings, \eg websites or services, exist~\cite{cedexis,dyn,xdn}, however, there is currently little to no understanding of their infrastructures, customers, and the effects on the global \ac{cdn} landscape.
With respect to Meta-\acp{cdn} and especially Cedexis, Xue \etal~\cite{xue2017cdn} are the first to provide brief performance figures about the selected \acp{cdn}, focusing on deployments in China.
We aim at more broadly characterizing Cedexis as a whole while looking at their infrastructure and performance on a global scale.
Nevertheless, we find, similar to Xue, partly suboptimal performance in terms of latency, yet, we acknowledge that routing decisions may have other goals than latency.
This argument is reinforced by Mukerjee \etal~\cite{mukerjee2016broker}, which is closest to our work. 
They analyze the effect of brokers, \ie \ac{cdn} selectors, on \acp{cdn}, characterize potential problems and propose a new interface between these brokers and \acp{cdn}.
While a closer interaction may improve certain aspects, it remains open whether a Meta-CDN such as Cedexis does, in fact, harm a \ac{cdn}'s profitability.
We cannot confirm a raised concern that a broker \emph{might} prefer certain \acp{cdn} in certain regions, as we find similar CDN choices worldwide.

The goal of this paper is to extend the currently limited understanding of Meta-CDN operation by characterizing Cedexis as a prominent example of a generic Meta-CDN.
Exemplified by understanding its overall deployment, customers, and the effects Cedexis, we aim to provide a first step towards a better understanding of Meta-CDNs in general.

\vspace{-1em}
\section{Characterizing a Meta-CDN}
\label{sec:meta_cdn}
\vspace{-1em}
The general motivation behind a Meta-CDN is to enable custom and dynamic routing of requests to content that is multi-homed in different \acp{cdi}.
A CDI can involve any infrastructure ranging from simple (cloud-hosted) servers to complex CDNs~\cite{poese2010padis}.
Multihoming content on different \acp{cdi} enables content providers to optimize for availability, performance, or operational costs.
By utilizing a Meta-CDN, content providers can realize {\em custom} routing logic to direct traffic to the available \acp{cdi}.
Such custom routing logic can be motivated by \acp{cdi} that offer better performance and/or lower costs in certain geographic regions, at certain times of the day, or request volumes.
We refer to an infrastructure that enables routing between CDIs with customer-provided routing logic as a {\em Meta-CDN}, a concept that is also referred to as Multi-CDN selector~\cite{xue2017cdn} and has similarities to auction-based CDN brokers~\cite{mukerjee2016broker}.
Since the individual routing approaches employed by content providers at the Meta-CDN are unknown to the involved \acp{cdi}, directed traffic and thus generated revenue gets harder to predict.
In particular, since decisions can be based on active performance measurements by the Meta-CDN, a (single) delivery of bad performance by the probed \ac{cdi} can result in rerouting traffic to a competing \ac{cdi} and thus losing revenue.
While the effects of Meta-CDN operation are relevant to Internet operation, little is known about Meta-CDNs.

To elucidate Meta-CDN operation, we start by characterizing Cedexis as a prominent example.
We base this characterization on showing
\begin{inparaenum}[\em i)]%
	\item its operational principles to select and routing between \acp{cdi} based on the Cedexis site (Section~\ref{sec:operation}) and 
	\item its current use in the Internet by analyzing its customer base in Section~\ref{sec:customer} based on our measurements.
\end{inparaenum}
Both perspectives provide a first understanding of the principle mechanisms with which Meta-CDNs can influence content distribution and their current deployment in the wild.

\subsection{Operation Principles}
\label{sec:operation}
\vspace{-0.5em}
\begin{figure}[t]
	\includegraphics[width=\columnwidth]{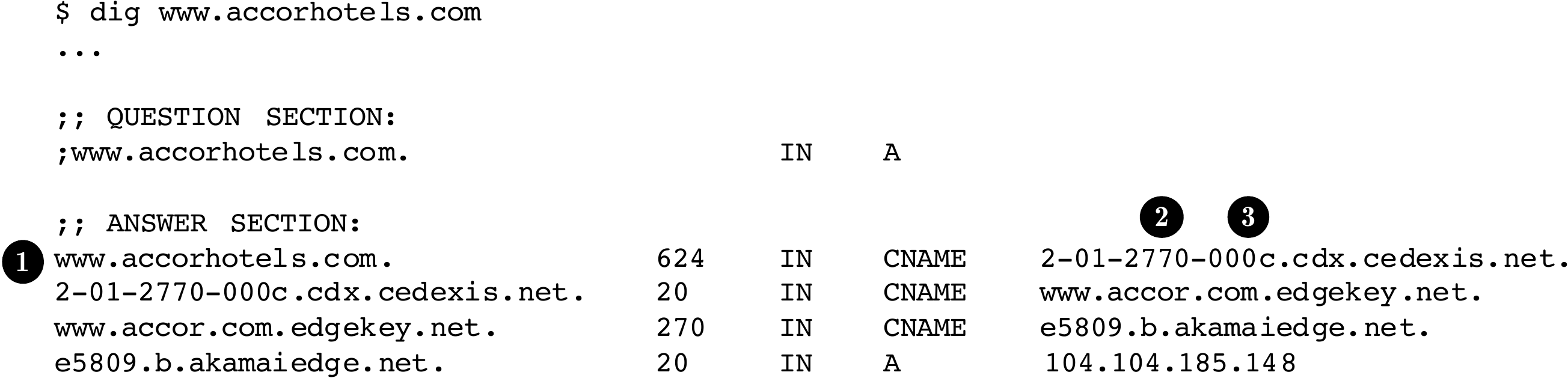}
	\vspace{-2em}
	\caption{Exemplary output of \texttt{dig} resolving a customer domain managed by Cedexis. 
	The requesting user is redirected by the Cedexis authoritative DNS with a CNAME to the \ac{cdn} selected to handle the request.
	The Cedexis CNAME containts the customer ID \protect\encircle{2} (C\textsubscript{ID}) and a per-customer application ID \protect\encircle{3} (App\textsubscript{ID}).}
	\label{fig:cedexis_dig}
	\vspace{-15pt}
\end{figure}

Like most \acp{cdn}, Cedexis employs a DNS-based redirection scheme, similar to Akamai~\cite{nygren2010akamai}, to redirect the requesting user to the \ac{cdi} selected for content delivery.
This redirection scheme is based on Canonical Name (CNAME) records which transfer the requesting user between the different authoritative name servers (NS).
We exemplify this scheme in Figure~\ref{fig:cedexis_dig}.
Starting at the original domain (\protect\encircle{1}), the user gets transferred to the Cedexis NS, which then selects a final \ac{cdi}.
The configured static Cedexis CNAME includes a Cedexis customer ID (C\textsubscript{ID}) (\protect\encircle{2}) and configuration specific (App\textsubscript{ID}) \protect\encircle{3}.
Both identifiers enable the Cedexis NS to perform customer-specific request routing, once the NS is contacted by the client's DNS resolver for name resolution.
Similar to classic \acp{cdn}, routing can be subject to a user's location, e.g., identified by DNS resolver IP or EDNS0 client subnet extension.
The Cedexis NS then points to the selected \ac{cdi}, which can be an IP address in an A resource record or another CNAME, \eg pointing to CDN (Akamai in this particular example).
The selected CDI can then repeat this process to select the final server handling the request or point to another CDI.
By realizing routing in the DNS, Cedexis redirects requesting users to a \ac{cdi} {\em before} a connection to the \ac{cdi} is established.
This way, it is not involved in the actual content delivery itself and thus does not alter the performance or security properties provided by the selected \ac{cdi}.
 
\afblock{CDI Selection Options.}
The above-stated request routing approach can be arbitrarily dynamic, \ie the user to \ac{cdi} mapping in the DNS can change at any time, only limited by the cacheability of their DNS records.
This aspect is utilized to enable content providers to realize custom routing logic within Cedexis using three components:
\begin{inparaenum}[\em i)]%
	\item \emph{Openmix} enables Cedexis customers to configure individual routing logic.
	Customers can choose between \texttt{optimal RTT} to select the \ac{cdi} with the lowest RTT to the requesting user, \texttt{round-robin} balancing between all \acp{cdi} configured by a customer, \texttt{throughput} to select the \ac{cdi} with the highest throughput, a \texttt{static routing} or by executing \texttt{customer provided code}.
	This routing logic can be informed by
	\item \emph{Radar}, a large database for decision making based on active browser-based \ac{cdn} measurements performed by website visitors, and
	\item \emph{Fusion}, to retrieve data from \acp{cdn}.
\end{inparaenum}
Every customer has the ability to configure site-specific behavior (\ie \emph{Apps}) which results in different App\textsubscript{IDs} in the DNS.
This way, a customer can configure different routing profiles for \textit{downloads.domain.tld} and for \textit{images.domain.tld.}
We next describe the two data sources and the Openmix platform to realize custom routing decisions.

\afblock{Radar.} Cedexis provides a community-driven \ac{cdn} performance measurement platform called Radar.
Radar employs active measurements performed within the web browser of visitors of Cedexis-managed websites.
The in-browser measurements require the Cedexis customers to embed a Javascript in their website.
Once visited, the web browser triggers the website's \texttt{onLoad} event, the embedded Javascript waits for a user configurable timeout (default \unit[2]{s}) and starts requesting probe instructions from Cedexis.
Users can configure private probes (\ie to estimate their own performance) and can choose to activate community probes (\ie enabling Cedexis to measure other infrastructures).
Probes can serve different purposes, latency or throughput measurements, which are expressed through a small \unit[43]{Byte} file (latency) or \unit[100]{kB} file (throughput) that is fetched via HTTP.
After having performed the measurements, the Javascript reports the obtained values back to Cedexis such that they can later be used to guide performance decisions in the CDN selection process and to inform site operators about their site's performance.

\afblock{Fusion.} While Radar enables to realize performance-based routing decisions, Fusion enables realize decisions on statistics directly from a user's \acp{cdn}.
\acp{cdn} typically offer statistics about traffic shares, quotas, budgets, performance and other KPIs in their web interfaces.
By accessing  these, Cedexis enables their customers to not only draw performance decisions but also business decisions (e.g., \emph{will I hit my quota soon?}).

\afblock{Openmix.} Both Radar and Fusion data are available to customers in the Openmix platform.
Openmix enables customers to customize the DNS resolution process by providing custom Javascript code that is executed in the DNS resolution step.
Within this code, customers can define their subsequent candidate \ac{cdi} choices and request measurement data (e.g., availability, latency, throughput) for these.
While the system is powered by many probes, only singles values are returned suggesting that Cedexis preprocesses the measurement data, yet we were unable to find more information on this process.
Thus, Openmix is used to realize customer-specific routing decision performed within the DNS resolution, i.e., directing traffic to a target \ac{cdi} via a DNS CNAME. 

\takeaway{Cedexis offers its customers to realize site-specific, fine-granular, and dynamic traffic routing to \acp{cdi}, \eg based on customer-provided code and informed by rich measurement data. The performed traffic routing is hard to predict (e.g., for \acp{cdi}).}

\vspace{-0.5em}
\subsection{Customers}
\label{sec:customer}
\vspace{-0.5em}
Before we analyze the infrastructure and configuration of Cedexis, we want to shed light on their customer base (in anonymous form).
We are interested in which companies and businesses leverage this additional service on top of classical \ac{cdn} infrastructures.

\begin{table}[t]
	\begin{subtable}[c]{0.8\textwidth}\centering
		{\begin{tabular}{r c}
     			\textbf{Type} 				& \textbf{Share} \\
			\Xhline{0.75pt}				
			Business					& 17.7\% \\
			\rowcolor{gray!25}IT			& 12.1\%\\ 
			News					& 11.3\%\\ 
			\rowcolor{gray!25}Gambling	& 11.3\%\\ 
			Shopping					& 8.1\%\\ 
			\rowcolor{gray!25}Gaming		& 8.1\%\\ 
   		\end{tabular}}
	\hspace{17.5pt}
		{\begin{tabular}{r c}
     			\textbf{Type} 				& \textbf{Share} \\
			\Xhline{0.75pt}
			Unknown					& 8.1\%\\ 
			\rowcolor{gray!25}Goods		& 5.6\%\\ 
			Automotive				& 5.6\%\\ 
			\rowcolor{gray!25}Advertising	& 3.2\%\\ 
			Streaming					& 2.4\%\\ 
			\rowcolor{gray!25}Television	& 1.6\%\\ 
   		\end{tabular}}
	\hspace{17.5pt}
		{\begin{tabular}{r c}
     			\textbf{Type} 				& \textbf{Share} \\
			\Xhline{0.75pt}
			Social					& 1.6\%\\ 
			\rowcolor{gray!25}CDN		& 1.6\%\\ 
			Patents					& 0.8\%\\ 
			\rowcolor{gray!25}Banking		& 0.8\%\\ 
			\\
			\\
   		\end{tabular}}
	\subcaption{Classification}
	\label{tab:customer:types}
	\end{subtable}%
	\begin{subtable}[c]{0.2\textwidth}\centering
		{\begin{tabular}{r c}
     			\textbf{Service} 				& \textbf{Share} \\
			\Xhline{0.75pt}					
			Web						& 62.7\%\\ 
			\rowcolor{gray!25}Unknown	& 15.6\%\\ 
			Assets					& 12.9\%\\ 
			\rowcolor{gray!25}Media		& 5.4\%\\ 
			API						& 2.3\%\\ 
			\rowcolor{gray!25}Bulkdata	& 1.1\%\\ 
   		\end{tabular}}
	\subcaption{Services}
	\label{tab:customer:services}
	\end{subtable}
\vspace{-1em}
\caption{Cedexis customer information obtained from manual inspection of websites served for different customer IDs.
Please note that customers may operate multiple services, \eg multiple brands of one holding company.}
\vspace{-3em}
\label{tab:customer}
\end{table}

\afblock{DNS Measurement Methodology.}
Our approach is twofold.
First, we leverage the encoded customer and application IDs in the CNAME structure (see step \protect\encircle{2} \& \protect\encircle{3} in Figure~\ref{fig:cedexis_dig}) to {\bf enumerate customers}.
Applications are used by customers to define different routing profiles that map to the available \ac{cdi}s, so e.g., a customer may have one profile that is optimized for latency, and another for throughput.
Conveniently, App IDs start at 1.
Thus our approach is to simply enumerate customers by resolving all {\texttt{2-01-(C\_ID)-(App\_ID).cdx.cedexis.net} domains.
As customer and application ID each have 4 hex characters, we would need to probe > \unit[2.5]{B} ($16^8$) domains.
To scale-down our DNS resolution, we only enumerate the first 256 application IDs for each customer, resulting in resolving roughly \unit[16]{M} domains.

\afblock{Domain Lists.}
Second, we probe domain lists to study the usage of the enumerated CNAMES in the wild and to discover application IDs beyond the enumerated 256 IDs.
We thus resolve the A of \texttt{domain.tld} and the A \texttt{www.domain.tld} records for all domains in the 
\begin{inparaenum}[\em i)]%
	\item .com/.net (obtained by Verisign),
	\item .org (obtained from PIR),
	\item .fi (obtained from Ficora), 
	\item .se/.nu (obtained from IIS),
	\item .new gTLD zones (obtained from ICANN's Centralized Zone Data Service),
	\item obtained from our passive DNS probe, and
	\item the Alexa Top 1M list.
\end{inparaenum}
We additionally include the Cisco Umbrella Top 1M list~\cite{umbrella}, which is based on the most frequent queries to OpenDNS resolvers and additionally contains subdomains, \eg \texttt{images.domain.tld}.
Depending on the size of the list, we perform daily or weekly resolutions for four weeks in August 2017 and extract all domains which have a CNAME pointer containing \texttt{*.cedexis.net}.

\afblock{Customer List.}
We combine both data sets to a customer list that will form the basis for probing Cedexis globally in Section~\ref{sec:globalview}.
The list contains all customer application tuples, of which \unit[84]{\%} were discovered in the enumeration step, \unit[11.2]{\%} were discovered in both the enumeration and in the domain lists, and \unit[4.8]{\%} solely in domain lists.
The reasons for the latter are application IDs larger than 256, which were not part of our enumeration.
Out of all customers, \CDXfirstMostAppCount{} (\CDXsecondMostAppCount{}) have only \CDXfirstMostAppsNumber{} (\CDXsecondMostAppsNumber{}) application(s) configured.
We also observe \CDXLargestAppCount{} customer having \CDXLargestAppsNumber{} configured.
By resolving the domain lists, we find 4609 (sub-)domains pointing to \unit[16]{\%} of all discovered (customer, application) tuples.
The remaining \unit[84]{\%} were not hit when resolving our domain lists.
For \unit[62.7]{\%} of all customer application IDs, we only find a single domain pointing to it.
We find 31 (6) application IDs managing more than 10 (100) domains, respectively.

\begin{figure}[t]
	\includegraphics[] {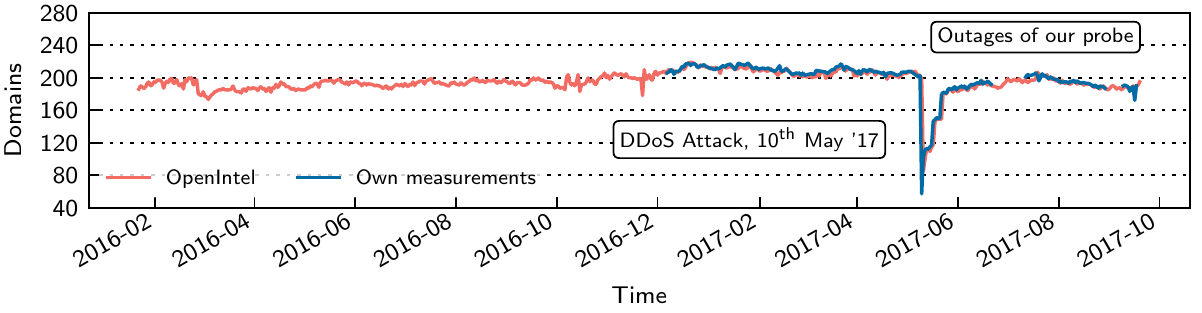}
	\vspace{-2em}
	\caption[]{Domains utilizing Cedexis in the Alexa 1M over time. The drop in May '17 was caused by a DDoS Attack on Cedexis~\cite{cedexis_ddos}. Unfortunately, the measurement probe located at our chair experienced two outages. However, the overlap of both scans motivates the further use of the OpenIntel data set.}
	\label{fig:cedexis_over_time}
	\vspace{-2em}
\end{figure}

We show the popularity of Cedexis over time among Alexa-listed domains in Figure~\ref{fig:cedexis_over_time}.
We set up our own regular DNS resolutions in December 2016 and further show regular Alexa resolutions performed by OpenINTEL~\cite{deji2016openintel} in the Netherlands for the same resource records.
First, we observe that both data sets overlap, suggesting that both are suitable for monitoring.
Minor fluctuations in the number of domains per day can mainly be attributed to fluctuations in the Alexa listed domains.
Second, we observe an outage of Cedexis in May 2017 which was caused by a DDoS attack on their infrastructure~\cite{cedexis_ddos}.
The outage motivated some customers to remove CNAME pointers to Cedexis in favor of pointing to operational \ac{cdn}s instead, causing a drop of > 120 domains in Figure~\ref{fig:cedexis_over_time}.

\afblock{Customer Classification.}
We next classify the discovered customers to highlight the variety of Cedexis customer' profiles.
To base this on an open classification scheme, we first tried to match customer domains against the Alexa Web Information Service API.
However, Alexa classifications exist only for 17\% of the queried domains and some classifications do not reflect the web pages' content.
To obtain a broader picture, we 
instructed a single human classifier to visit each web site and categorize it according to an evolving set of categories.
We show the resulting categorized web site content in Table~\ref{tab:customer}(a).
The table shows that Cedexis is used by a broad range of customers.
We further classify the used service in Table~\ref{tab:customer}(b).
The table shows that most customers use Cedexis for general web content delivery.
This includes few but large bulk download services, \eg \texttt{www.download.windowsupdate.com}.
This is in contrast to, \eg Conviva which is dedicated to video delivery.

\takeaway{Cedexis is utilized by a number of (large) web services. Decisions taken by Cedexis have the potential to impact larger bulks of Internet traffic.}

\vspace{-1em}
\section{A Global View on Cedexis}
\label{sec:globalview}
As Cedexis customers can realize routing decisions based on (network) location, we next take a global view on its customers by using globally distributed active measurements.

\afblock{Measurement Setup.} 
We base our measurements on 35 PlanetLab nodes located in 8 countries, 6 custom \ac{rpi} probes in 6 distinct German ISPs, and RIPE Atlas probes.
We chose PlanetLab and custom \acp{rpi} in addition to RIPE Atlas since they enable us to deploy custom software to perform frequent DNS resolutions.
As we do not include PlanetLab nodes located in Germany in our set, we refer to our deployed \acp{rpi} when mentioning \emph{DE} in figures or plots.
We selected only few PlanetLab nodes with high availability to repeatedly measure always from the same vantage points.
For our measurement, we instruct the Planet Lab and our \ac{rpi} nodes to resolve each domain every \unit[15]{min} and subsequently measure the latency to the resulting IPs.
Moreover, we keep track of all CNAME redirections to \acp{cdn} that we observe over the course of the measurement and also resolve these.
This way, we learn the set of configured \acp{cdn} for every probed domain.

\subsection{Infrastructure}
\label{sec:infrastructure}
\vspace{-0.5em}
\begin{figure}[t]%
	\captionsetup[subfigure]{width=0.9\columnwidth}
	\centering
	\hfill
	\begin{subfigure}[t]{.425\columnwidth}
	\includegraphics[width=\columnwidth]{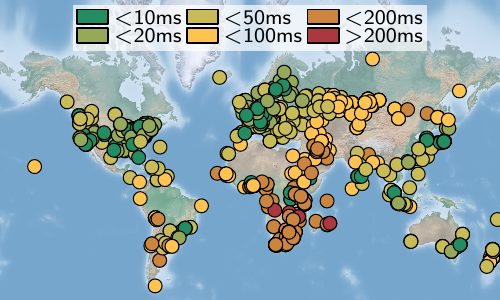}%
		\caption{Minimum \texttt{ping} RTTs to the four Cedexis authoritative nameservers.}
		\label{fig:cedexis:arch_deployment}%
	\end{subfigure}\hfill%
	\begin{subfigure}[t]{.425\columnwidth}
	\includegraphics[width=\columnwidth]{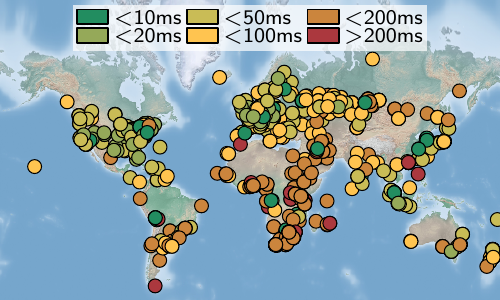}%
		\caption{Median DNS query times to resolve an \texttt{A} record via Cedexis}%
		\label{fig:cedexis:dns_overhead}%
	\end{subfigure}\hfill%
	\vspace{-.5em}
	\caption{RTTs and DNS Query times obtained from $\approx 870$ responsive RIPE Atlas Probes performing pings and DNS A Record requests.}
	\label{fig:cedexis:overall}
	\vspace{-15pt}
\end{figure}

\afblock{Authoritative DNS Deployment.}
Cedexis core functionality is based on a distributed infrastructure of authoritative name servers managing \texttt{*.cedexis.net}.
We find four servers configured in the DNS in our measurements and in the \texttt{.net} zone file.
We remark that a larger number exists which we found by enumerating their naming pattern.
However, they currently appear to be unused, i.e., not included in the \textit{.net} zone and are not discovered by our active DNS measurements.

To obtain a better understanding of its DNS infrastructure, we measure the ICMP echo (ping) latency to their authoritative name servers from $\approx 870$ responsive (out of 1000 selected) RIPE Atlas probes.
We repeated this measurement 30~times using the same set of probes and show the minimum RTT in Figure~\ref{fig:cedexis:arch_deployment}.
Based on these latency figures, we infer that Cedexis operates DNS servers located in North-America, Europe, and (probably) Asia and South America.
By analyzing individual latencies and manual traceroutes per server-IP (not shown), we observe latencies of <~\unit[10]{ms} from multiple regions (e.g., US-East, US-West, Europe, Hongkong) to the {\em same} DNS server IP.
Since these low latencies are lower than required by the speed of light between the respective regions, it suggests that the probed server-IPs are served using \emph{anycast} routing.

Since the additional indirection step through Cedexis contributes latency, we next measure the DNS resolution time of only their authoritative name servers.
In this step, we resolve the domain of a Cedexis customer from all authoritative name servers from the same RIPE Atlas probes, again repeated 30 times.
To limit the resolution to only involve the Cedexis DNS, we chose a customer domain which directly returns an A record instead of redirecting to another \ac{cdn}.
This is especially important if the lifetime of the DNS records is short (which we analyze in Section~\ref{sec:customerconfig}) and clients need to frequently contact the Cedexis DNS over and over again.
We show the median DNS query time in Figure~\ref{fig:cedexis:dns_overhead} and observe that the DNS query times follow the previously measured ping latencies.
Nevertheless, we observe few regions in which Cedexis appears to be uncovered as they involve high DNS resolution latencies, \eg Latin America or Africa.

\begin{figure}[t]%
	\captionsetup[subfigure]{width=0.9\columnwidth}
	\centering
	\hfill
	\begin{subfigure}[t]{.425\columnwidth}
	\includegraphics[width=\columnwidth]{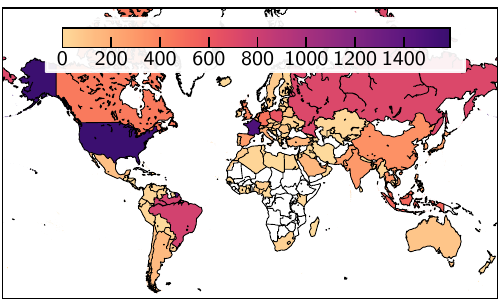}%
		\caption{Locations of confirmed events.}
		\label{fig:cedexis:event}%
	\end{subfigure}\hfill%
	\begin{subfigure}[t]{.425\columnwidth}
	\includegraphics[width=\columnwidth]{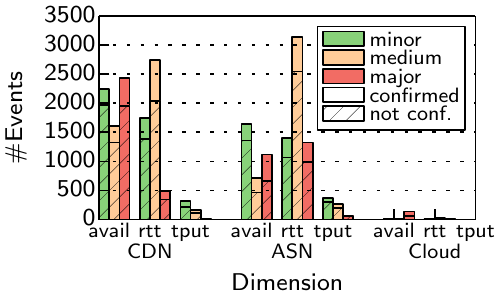}%
		\caption{Event types, status, and severity.}
		\label{fig:cedexis:event_types}%
	\end{subfigure}\hfill%
	\vspace{-.5em}
	\caption{Cedexis events reported from 9th October 2017 to 8th January 2018.}
	\label{fig:cedexis:events_overall}
	\vspace{-15pt}
\end{figure}

\afblock{Radar Community Probes.}
Cedexis customers can realize routing decisions that are based on active measurements of current service performance probed by visitors of Cedexis managed websites (Radar platform, see Section~\ref{sec:operation}).
Understanding this data is interesting since it can influence routing decisions.
As the Radar data is not publicly available, we instead analyze live feeds of network events detected by Radar and published at \url{https://live.cedexis.com}.
The events report three classes of metrics: \emph{latency}, \emph{throughput}, and \emph{availability} for \emph{\ac{isp}}, \emph{\ac{cdn}}, and \emph{cloud} infrastructures.
An event can be a latency in- or decrease, an outage, or a change in throughput.
A \ac{cdn}/cloud event is detected if it was reported by visitors from 5 different ASes.
Likewise, an AS event is detected if it concerns 5 \ac{cdn}s or clouds (see \url{live.cedexis.com}).
Each event can be classified by severity into \emph{minor}, \emph{medium}, and \emph{major}.
Apart from being used in their decision-making process, this data allows monitoring the reported infrastructures.
We thus monitored the data feed from October 9, 2017, to January 8, 2018.

Reported events further provide information {\em where} visitors of Cedexis-managed sites are located.
This is based on fact that Cedexis customers embed Javascript measurement code into their sites that reports performance figures to the Radar platform.
While the number of events is likely uncorrelated with the number of website visitors, it at least indicates the presence of a visitor in the reported AS or country.
Therefore, we show the distribution of the number of events per-country in Figure~\ref{fig:cedexis:event}.
We observe almost no events in Africa, suggesting that Cedexis customers do not have a large user base in Africa, which also coincides with the suboptimal DNS deployment there.
While we see events in almost every country, most events are reported in Central Europe, North America, Brazil, and Russia.

We next analyze the reported events by their type, shown in Figure~\ref{fig:cedexis:event_types}.
The figure shows the number of events per event type categorized to availability (avail), latency (rtt), and throughput (tput) for \ac{cdn}s, AS, and cloud providers.
Every bar is divided in the amount of confirmed and unconfirmed events.
We observed that an event is marked as \emph{confirmed} when it was reported for at least 9 minutes and the rolling variance of measurements from the last 5 hours exceeds an event and severity level specific threshold:
\eg a latency increase of \unit[100]{\%} - \unit[200]{\%} for a \ac{cdn} is considered as minor, while an increase between \unit[200]{\%} and \unit[500]{\%} is considered as medium severity.
We find most of the reported events to concern \ac{cdn}, followed by ASes.
The high amount of major availability events concern CacheFly \ac{cdn} outages during our measurement period.

\takeaway{We observe visitors of Cedexis-managed sited from almost every country. Yet, its anycast DNS platform is suggested to be based in the US, Europe, and Asia. Users in other countries can be subject to higher DNS query latencies.}

\subsection{How Customers utilize Cedexis}
\label{sec:customerconfig}
\vspace{-0.5em}

\begin{figure}[t]%
	\center
	\begin{subfigure}[b]{.40\columnwidth}
	\includegraphics{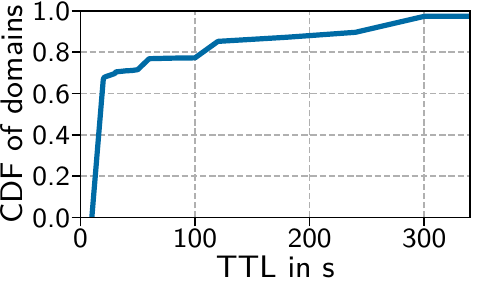}%
		\vspace{-0.5em}
		\caption{Customer configured TTL\\from Cedexis to the next CDN.}
		\label{fig:cedexis:ttl_first}%
	\end{subfigure}\hfill%
	\begin{subfigure}[b]{.60\columnwidth}%
	\resizebox{\linewidth}{!}{%
	\renewcommand{\arraystretch}{.6}
		\newcolumntype{P}[1]{>{\centering\arraybackslash}p{#1}}%
	\begin{tabular}[t]{P{0.8cm} P{1.5cm} p{1.8cm} c}%
     			\textbf{Rank} 				& \textbf{Used by}    & \textbf{CDN}    & \textbf{TTL in seconds} \\
			\Xhline{0.75pt}
			\TTLRankAkamai	&	\TTLRankChosenTimesAkamai	&	\TTLRankNameAkamai	&	\vspace{1pt}\includegraphics{\TTLRankBoxFilepathAkamai}  \\
			\TTLRankCDNetworks	&	\TTLRankChosenTimesCDNetworks	&	\TTLRankNameCDNetworks	&	\vspace{1pt}\includegraphics{\TTLRankBoxFilepathCDNetworks}  \\
			\TTLRankDNSDD	&	\TTLRankChosenTimesDNSDD	&	\TTLRankNameDNSDD	&	\vspace{1pt}\includegraphics{\TTLRankBoxFilepathDNSDD}  \\
			\TTLRankEdgecast	&	\TTLRankChosenTimesEdgecast	&	\TTLRankNameEdgecast	&	\vspace{1pt}\includegraphics{\TTLRankBoxFilepathEdgecast}  \\
			\TTLRankLevel3	&	\TTLRankChosenTimesLevel3	&	\TTLRankNameLevel3	&	\vspace{1pt}\includegraphics{\TTLRankBoxFilepathLevel3}  \\
			\TTLRankCloudflare	&	\TTLRankChosenTimesCloudflare	&	\TTLRankNameCloudflare	&	\vspace{1pt}\includegraphics{\TTLRankBoxFilepathCloudflare}  \\
			\TTLRankCDNWD	&	\TTLRankChosenTimesCDNWD	&	\TTLRankNameCDNWD	&	\vspace{1pt}\includegraphics{\TTLRankBoxFilepathCDNWD}  \\
			\TTLRankChinaCache	&	\TTLRankChosenTimesChinaCache	&	\TTLRankNameChinaCache	&	\vspace{1pt}\includegraphics{\TTLRankBoxFilepathChinaCache}  \\
			\TTLRankCloudfront	&	\TTLRankChosenTimesCloudfront	&	\TTLRankNameCloudfront	&	\vspace{1pt}\includegraphics{\TTLRankBoxFilepathCloudfront}  \\
\vspace{-1.52em} \TTLRankHighwinds	&	\vspace{-1.6em}\TTLRankChosenTimesHighwinds	&  \vspace{-1.6em} 	\TTLRankNameHighwinds	&	\vspace{0pt} \includegraphics{\TTLRankBoxFilepathHighwinds}  \\
   	\end{tabular}
	}
	\vspace{-0.5em}
		\caption{TTLs of \texttt{A}-record for top 10 used CDNs. (Note the gap in the time scale to display Edgecast using anycast.)}%
		\label{fig:cedexis:dns_per_cdn}%
	\end{subfigure}\hfill%
	\vspace{-0.5em}
	\caption{DNS TTLs experienced among Cedexis-enabled domains. For a) mappings from Cedexis to the subsequent entry and b) the CDNs used for the final delivery.}
	\label{fig:cedexis:ttls}
	\vspace{-2em}
\end{figure}

\afblock{DNS TTL.}
The DNS Time To Live (TTL) defines the time a record can be cached by DNS resolver and thus the timespan between Cedexis balancing decisions.
A small TTL allows more frequent switches at the cost of more frequent DNS queries to the Cedexis DNS infrastructure.
This query latency can be significant,  depending on the DNS resolver location.

Figure~\ref{fig:cedexis:ttl_first} depicts the CDF of the TTLs for the validity of the CNAME-mappings from Cedexis to the subsequent entity (see 2\textsuperscript{nd} CNAME in Figure~\ref{fig:cedexis_dig}) for all customer domains.
We did not observe country-specific settings.
Around 67\% of all domains have configured a TTL of at most \unit[20]{s}, indicating a rather short time scale enabling rapid reactions to changes.
The next 30\% are within \unit[300]{s}, denoting an already moderate reaction time while around 3\% have configured higher TTLs.
Higher TTLs {\em can} hint to non-latency-based, but rather throughput or cost-based optimizations.

To compare these configurations to TTLs deployed by CDNs, we show the \texttt{A}-record TTLs for the top 10 \acp{cdn} in Figure~\ref{fig:cedexis:dns_per_cdn}.
To the right of every \ac{cdn}, the figure shows the boxplot of TTLs observed for the \texttt{A}-records of all resolutions we performed.
We see that the top 3 \acp{cdn} use a short TTL in the range of most Cedexis CNAMEs, whereas Edgecast has a lifetime of one hour (probably due to their use of anycast).

\afblock{DNS Resolution Time.}
When employing Cedexis, an additional step in DNS resolution is required to enable \ac{cdn} balancing.
Figure~\ref{fig:cedexis:cdn_lookup_latency} compares the latency for resolving (from our Planet Lab sites) a mapping at Cedexis, in case of multi-staged \acp{cdn} a further CNAME redirect (CDN) and the final resolution of the A-record.
We observe that Cedexis performs similarly to the other \acp{cdn}.
However, while this hints at a good DNS deployment for our vantage points, it also means that using Cedexis inflates the latency of a DNS lookup. 
Given the on average short TTLs, users will often incur an additional added latency when using Cedexis-enabled websites.

\afblock{CDN usage of Customers.}
We next check how many \acp{cdn} are being used by Cedexis customers.
Figure~\ref{fig:cedexis:cdn_geo_choice} illustrates the overall \ac{cdn} selection frequency for every PlanetLab and \ac{rpi} node location over the period of one month for all discovered Cedexis domains.
We find that domains are usually only using one or two CDNs while there are few that use more.
This finding is consistent between geolocations.
Only when looking at the how often which CDN is actually selected (note: Figure~\ref{fig:cedexis:dns_per_cdn} did only show the share of domains using the CDN) we see a small geographic difference in China (CHN).
Here ChinaCache is selected more often than in other geolocations, nevertheless, apart from this, all geolocations behave similarly.
This finding contrasts a finding on the Conviva network~\cite{mukerjee2016broker} showing a bias in which some \ac{cdn}s are selected more often than others in specific countries.
In summary, we do not observe that Cedexis customer set country-specific routing decisions.

\begin{figure}[t]%
\begin{subfigure}[t]{.49\columnwidth}
	\includegraphics[width=\textwidth]{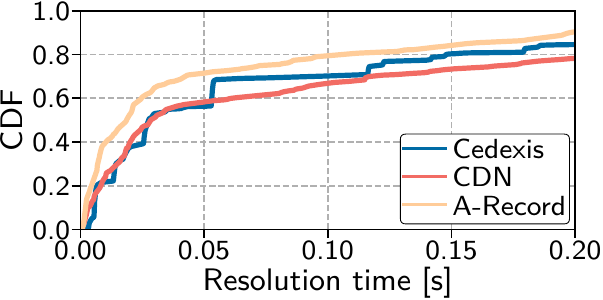}%
		\caption{Lookup latency.}
		\label{fig:cedexis:cdn_lookup_latency}%
\end{subfigure}\hfill%
	\begin{subfigure}[t]{.49\columnwidth}
	\includegraphics{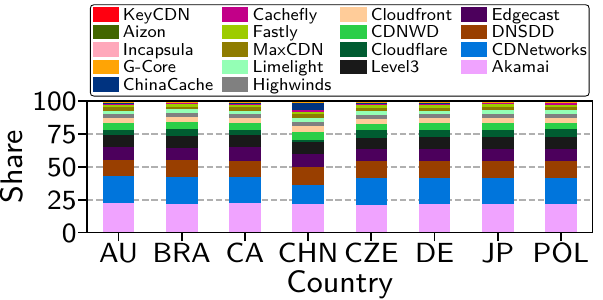}%
		\caption{Share of CDN choices per country.}
		\label{fig:cedexis:cdn_geo_choice}%
		\end{subfigure}
	\vspace{-1em}
	\caption{Lookup latency in DNS resolution and final CDN choices of Cedexis customers.}
	\vspace{-2em}
\end{figure}

To extend our global view beyond the PlanetLab nodes, we next resolve the discovered Cedexis domains from open DNS resolvers obtained from \url{public-dns.info}.
To avoid overloading (low-power) devices on user-premises (\eg home routers), we exclude all resolvers whose DNS names indicate access lines (\eg ``pppoe'', ``dial-up'', or ``dsl'').
We further only select resolvers with an availability $>89\%$.
In total, this leaves us with 1998 resolvers in 161 countries, out of which 67 \emph{never} successfully responded.
We resolve all Cedexis customer domains using all resolvers every two hours for four days.
Subsequently, we group the reported results by continent and compare the top selected \ac{cdn}.
We observe that \unit[66.9]{\%} always chose the same \ac{cdn} in every continent.
For the remaining, we observe disagreement, \ie different \ac{cdn} are chosen on each continent: \unit[30.4]{\%} have two and \unit[2.7]{\%} three \ac{cdn}s present.
We compare the complete \ac{cdn} choices in countries of our PlanetLab nodes to their mapping results and observe  similar distributions as in Figure~\ref{fig:cedexis:cdn_geo_choice} (not shown).

\takeaway{Most Cedexis domains configure short TTLs to enable frequent switches. We observe that most domains indeed balance between few \acp{cdn}.
Switches pose a challenge to each \ac{cdn} since traffic gets harder to predict.}

\vspace{-0.5em}
\subsection{Latency Perspective}
\vspace{-0.5em}
We next take a latency perspective on Cedexis choices, \ie comparing the latency of the chosen \ac{cdn} to all configured \acp{cdn} for every customer domain.
Thus, we measure the latency to every \ac{cdn} IP by performing ICMP pings.
We chose ICMP pings over more realistic HTTP requests since the pings do not generate accountable costs for the probed customers but remark that the ping latency can differ from actual HTTP response latencies.
Yet, the ping latency can be a reasonable performance and distance indicator for the different \ac{cdn} caches.

Figure~\ref{fig:cedexis:ping_diff_rel} shows the relative latency inflation for cases where Cedexis did not choose the latency-optimal CDN.
We observe that around 50\% of all resolutions are only marginally worse than the optimal choice regardless of the geographic location (selection shown).
The curves than start to flatten indicating increased latency inflation.
We observe two groups, where around 10\% (20\%) of the choices exhibit a latency increase of over 250\%.
The observed increase can be in the range of few ms for nearby servers that would still deliver good performance.
Therefore, we show the absolute difference of all sub-optimal decisions in Figure~\ref{fig:cedexis:ping_diff_abs}.
We observe that $\approx$ 50\% of all decisions indeed only differ in a couple of milliseconds, indicating almost no noticeable difference.
Apart from the nodes in Brazil and Japan (not shown), around 90\% of all choices are even within a \unit[20]{ms} difference.

\begin{figure}[t]%
\begin{subfigure}[b]{.49\columnwidth}
	\includegraphics{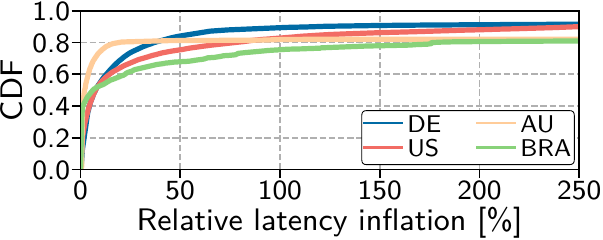}
	\caption{Relative difference to optimal choice.}
	\label{fig:cedexis:ping_diff_rel}%
\end{subfigure}\hfill%
\begin{subfigure}[b]{.49\columnwidth}
	\includegraphics{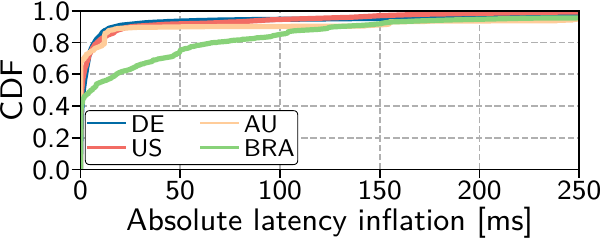}
	\caption{Absolute difference to optimal choice.}
	\label{fig:cedexis:ping_diff_abs}%
\end{subfigure}
\vspace{-.5em}
\caption{Ping difference to the \emph{fastest} CDN (RTT) when the choice was not optimal.}
\vspace{-15pt}
\end{figure}

We remark that it is difficult to assess the quality and correctness of CDN choices as we do not know the routing metrics that are employed by Cedexis customers.
Our measurements are motivated from the perspective of an end-user, who is interested in performance, not in potentially monetary business decisions. 

\takeaway{All available CDNs would deliver good latency figures in most tested cases, suggesting that the choice of \ac{cdn} performed by Cedexis would not largely impact end-user experience.}

\vspace{-1em}
\section{Conclusion}
\vspace{-1em}
This paper presents a broad assessment of a Meta-CDN deployment, exemplified by dissecting Cedexis as a representative generic platform.
By this, we enrich the existing literature describing the Meta-CDN concept with an empirical assessment of a large operator.
We shed light on Cedexis customers, technology, and performance.
Cedexis DNS deployment, even though using anycast, appears to be focussing on Europe, North America and parts of Asia indicated by high latencies in other regions.
We find customers to mostly configure short TTL values enabling fast reactions and we indeed observe that most domains balance between few \acp{cdn}.
By assessing ping latencies to all available CDNs, we observe that most available CDNs offer good performance from our distributed probe platform.
However, we also find a range of sub-optimal latency choices which can indicate routing metrics other than latency.

These unpredictable routing decisions implemented by customers using a Meta-CDN pose a challenge to CDN operators since inbound traffic gets much harder to predict.
In particular routing decisions can be based on active measurements by the Meta-CDN---thus, bad performance can result in rerouting traffic and thus losing revenue.
Studying Meta-CDNs and their consequences thus pose an interesting angle for future work.

\vspace{-1em}
\section*{Acknowledgments}
\vspace{-0.75em}
This work has been funded by the DFG as part of the CRC 1053 MAKI.
We would like to thank Dean Robinson (Univ.\ Michigan) for his early contributions to the Alexa analysis, our shepherd Ignacio Castro and the anonymous reviewers.
\vspace{-1.0em}
\bibliographystyle{abbrv}
\bibliography{literature_short}
\end{document}